\documentclass{ws-procs9x6}

\newcommand{\refeq}[1]{(\ref{#1})}

\begin{document}

\title{THE INFLUENCE OF LORENTZ VIOLATION \\
ON UHE PHOTON DETECTION}

\author{G.I.\ RUBTSOV, P.S.\ SATUNIN\footnote{Corresponding
    author, e-mail: satunin@ms2.inr.ac.ru}, 
and S.M.\ SIBIRYAKOV}

\address{Institute for Nuclear Research of the Russian Academy of Sciences,\\
 60th October Anniversary Prospect, 7a, 117312  Moscow, Russia}

\begin{abstract}
We show that violation of the Lorentz symmetry in quantum
electrodynamics can suppress the rates of the
interactions crucial for the formation of 
photon-induced air showers, such as pair
production on nuclei and in the geomagnetic field. As a consequence,
the allowed region in the space of Lorentz-violating parameters will
be seriously 
restricted if several photons with energies $\gtrsim\,10^{19}\,
\mbox{eV}$ are detected. 
\end{abstract}

\bodymatter

\section{Introduction}

The presence of a photon component in the ultra-high-energy (UHE)
cosmic rays is an important open question in astroparticle physics.
An answer to it will give information on astrophysical issues such
as the composition and origin of the UHE cosmic rays, as well as on
the fundamental topic of possible modifications of the space-time
symmetries\cite{Galaverni:2008yj}. 

In the standard physical picture, a primary UHE photon reaching the
Earth interacts in the atmosphere and produces an extensive air shower
of particles with lower energies that can be detected by the
ground-based experiments. The characteristics of the shower are
sensitive to the altitude, at which the first interaction giving rise
to the shower takes place. This, in its turn, is determined by
the cross section of the first interaction. 
At energies $\sim 10^{19}\,\mbox{eV}$
the dominant channel of the first interaction is electron-positron 
pair production on nuclei in the
atmosphere, the so-called 'Bethe-Heitler process'\cite{Bethe:1934za}. At
higher energy ($10^{20}\,\mbox{eV}$ and above) the pair production in
the geomagnetic field becomes important leading to the formation of a
preshower above the atmosphere. The photon showers initiated by these
processes can be identified by the cosmic
ray detectors using various observables.

The cross section of the first interaction, and consequently the
characteristics of the shower, may be affected by new physics.
In this paper we study the influence of possible violation of the 
Lorentz invariance (or Lorentz violation (LV) for short)
on the aforementioned processes, and the bounds on LV
parameters that can be established if the UHE photons are detected. 

\section{The model}

To make the quantitative predictions we study the model of LV 
quantum electrodynamics with the following Lagrangian:  
\begin{eqnarray}
&\mathcal{L}=\bar{\psi}\left( i\gamma^\mu D_\mu - m \right) \psi -\frac{1}{4}F_{\mu\nu}F^{\mu\nu} 
+\notag\\+i\varkappa \bar{\psi} &\gamma^i D_i\psi + \frac{ig}{M^2}D_j\bar{\psi}\gamma^iD_iD_j\psi+ \frac{\xi}{4M^2} F_{kj}\partial_i^2 F^{kj}, 
\label{Lagr} 
\end{eqnarray} 
where $\varkappa,\,g$ and $\xi$ are dimensionless parameters, the 
covariant derivative $D_\mu$ is defined in the standard way, 
$D_\mu\psi=(\partial_\mu+ieA_\mu)\psi$, and $M$ is the Planck mass. 
Summation over repeated indices with Minkowski metrics is
understood. This Lagrangian contains all inequivalent operators
of dimension up to 6 that are  
rotationally invariant in the
preferred frame, gauge invariant, and CPT- and P-even. The 
requirements on the theory and their motivation are discussed in
detail in Ref. \refcite{Rubtsov:2012kb}.

From the Lagrangian \refeq{Lagr} one obtains the dispersion relations for photons
and electrons/positrons, 
\begin{equation}
E^2_\gamma=k^2+\frac{\xi k^4}{M^2}, \qquad \qquad   E^2_e=m^2+p^2\left( 1 + 2\varkappa \right) + \frac{2gp^4}{M^2}.
\end{equation}
Also the Feynman rules for the model \refeq{Lagr} are modified
compared to the Lorentz invariant case\cite{Rubtsov:2012kb}. 
For processes with an electron-positron pair in the final state one 
introduces the combination 
\begin{equation}
\omega_{\mathrm{LV}}(x)=-\varkappa k  - \frac{gk^3}{4M^2}(1+3x^2)
+\frac{\xi k^3}{2M^2}\;,
\end{equation}
which characterizes the energy transfer in the process.
Here $x\in[-1.1]$ denotes the asymmetry between the momenta of the
produced electron and positron (see Ref. \refcite{Rubtsov:2012kb} for the
precise definition). 
If $\omega_{\mathrm{LV}}(x)$ is larger than $2m^2/k$ for some $x$, vacuum
photon decay becomes
kinematically allowed. UHE photons decay very quickly into
electron-positron pairs and do not reach the Earth. Below
we restrict to the values of $\omega_{\mathrm{LV}}$ outside this range.

\section{Pair production on a nuclei}

Let us consider the Bethe-Heitler process. The standard result for 
the cross section reads\cite{Bethe:1934za} 
$\sigma_{\mathrm{BH}}=28Z^2\alpha^3/9m^2\log \left(183/Z^{1/3}\right)$, 
where $m$ is the electron mass,
$\alpha$ is the fine structure constant and $Z$ is the nucleus 
charge. Lorentz violation significantly suppresses the cross section
for negative $\omega_{\mathrm{LV}}(x)<-m^2/k$. 
For example, in the case $1 \ll k|\omega_{\mathrm{LV}}|/m^2\ll
\alpha^{-4} Z^{-4/3}$ the modified cross section takes the
form\footnote{Unlike the standard case, the cross
section is saturated at the maximal asymmetry between the momenta of
the pair, $x=\pm 1$.}\cite{Rubtsov:2012kb} 
\begin{equation}
\sigma_{\mathrm{BH}}^{\mathrm{LV}} \simeq  \frac{8Z^2\alpha^3}{3k|\omega_{\mathrm{LV}}(1)|}\log\frac{1}{\alpha Z^{1/3}}\cdot \log\frac{k|\omega_{\mathrm{LV}}(1)|}{m^2}.
\end{equation}
This expression is suppressed roughly by a factor
$m^2/k\left|\omega_{\mathrm{LV}}(1)\right|$ with respect to the LI result.

A future UHE photon detection by cosmic ray experiments would imply
that the cross section of the first interaction is not too much
suppressed compared to the standard expectation: otherwise the photon
would go through the atmosphere without developing a
shower. Conservatively, 
we will assume that the cross section does not differ by
more that an order of magnitude. This gives the bound 
$|\omega_{\mathrm{LV}}|<10 m^2/k$ for negative
$\omega_{\mathrm{LV}}$. Similar (actually, even stronger) bound for
positive $\omega_\mathrm{LV}$ follows from the absence of the vacuum
photon decay. Barring accidental cancellations we find that a
prospective detection of photons with energies 
$k\sim 10^{19}\,\mbox{eV}$ will allow to constrain the LV parameters
at the level
\begin{equation}
|\varkappa| \lesssim 10^{-25}~;~~~~ |g|, |\xi| \lesssim 10^{-7}\;.
\end{equation}
It is worth stressing that these bounds will be insensitive to any
astrophysical assumptions as to the origin and propagation of cosmic
rays through the interstellar medium.

\section{Photon decay in magnetic field}

Next we turn to the photon decay in geomagnetic field. We
use\cite{Satunin} the method of 'worldline instantons', proposed by
Affleck, Alvarez and Manton\cite{Affleck:1981bma} to study 
the Schwinger effect. The photon decay in a weak magnetic field is
represented as a tunneling process whose width is given by a
statistical sum of a certain auxiliary quantum-mechanical
system. This statistical sum can be evaluated in the saddle-point
approximation and yields the following exponential behavior for the
photon decay width,
\begin{equation}
\Gamma \propto \exp\left[-\frac{8m^3}{3k eH\sin\varphi}\left( 1 - \frac{k\cdot\omega_{\mathrm{LV}}(0)}{2m^2}\right)^{3/2}\right].
\end{equation}
Here $H$ denotes the magnetic field, $\varphi$ is an angle between the
photon momentum and the magnetic field. In the Lorentz invariant case,
$\omega_{\mathrm{LV}}(0)=0$, this answer coincides with the standard
one \cite{Klepikov}. We see that the process is exponentially
suppressed till the photon energy reaches $k\sim 8m^3/3eH\sin\varphi$,
which for the Earth magnetic field and $\sin\varphi\sim 1$ gives
$k\sim 10^{20}\,\mbox{eV}$. In the presence of LV with
$\omega_\mathrm{LV}(0)<0$ the exponent is enhanced and the effective
threshold energy is shifted upwards. Requiring that the shift is not
too large again implies $|\omega_{\mathrm{LV}}|\lesssim m^2/k$.
We conclude that a detection of
UHE photons at $k\sim 10^{20}\,\mbox{eV}$
with the preshower signature will yield the
constraints,\footnote{The bound on the positive $\omega_\mathrm{LV}$ again
  follows from the absence of vacuum photon decay.}
\begin{equation}
|\varkappa|<10^{-27}~;~~~~  |g|, |\xi| <10^{-11}\;.
\end{equation}

\section{Conclusion}

We have shown that a future UHE photon detection can establish very
strong two-sided bounds on the parameters describing deviations from LI
in quantum electrodynamics. These bounds will be insensitive to any
astrophysical assumptions about the origin of the UHE photons and
their propagation towards the Earth.

\section*{Acknowledgments}

This work has been supported in part by
the Grants of the Russian Ministry of Education and Science 8412 and
14.B37.21.0457,
the Grants of the 
President of Russian Federation NS-5590.2012.2, MK-1170.2013 
and by the RFBR grants
11-02-01528, 12-02-01203, 12-02-91323.


\begin{thebibliography}{xx}

\bibitem{Galaverni:2008yj}
  M.~Galaverni and G.~Sigl,
  Phys.\ Rev.\ D {\bf 78} (2008) 063003.

\bibitem{Bethe:1934za} 
  H.~Bethe and W.~Heitler,
  Proc.\ Roy.\ Soc.\ Lond.\ A {\bf 146} (1934) 83.

\bibitem{Rubtsov:2012kb}
  G.~Rubtsov, P.~Satunin and S.~Sibiryakov,
  Phys.\ Rev.\ D {\bf 86} (2012) 085012.

\bibitem{Satunin}
P.~Satunin, Phys.\ Rev.\ D {\bf 87} (2013) 105015.

\bibitem{Affleck:1981bma}
  I.~K.~Affleck, O.~Alvarez and N.~S.~Manton,
  Nucl.\ Phys.\ B {\bf 197} (1982) 509.

\bibitem{Klepikov}
N.~P.~Klepikov, Zh. Exp. and Theor. Phys., {\bf 26}, 19 (1954).



\end{thebibliography}
\end{document}